\documentclass[twocolumn,twoside,slac_two,nofootinbib,nobibnotes]{revtex4}
\usepackage{graphicx}
\usepackage{fancyhdr}
\begin{document}

\title{AGN Physics with the Cherenkov Telescope Array}

%

\author{A. Zech (for the CTA Consortium)}
\affiliation{LUTH, Observatoire de Paris, CNRS, Universit\'e Paris Diderot ; 5 Place Jules Janssen, 92190 Meudon, France}

\begin{abstract}
The Cherenkov Telescope Array (CTA), currently in its Preparatory Phase, will be the first 
open observatory for very high energy $\gamma$-rays from galactic and extragalactic sources. 
The international consortium behind CTA is preparing the construction of two large arrays of Cherenkov 
telescopes in the Northern and Southern Hemispheres with a performance that will be significantly 
improved compared to the current generation of arrays.

Its increased sensitivity and energy range will give CTA access to a large
population of Active Galactic Nuclei (AGN) not yet detected at very high energies and provide 
much more details on known TeV sources. While the low end of the CTA energy 
coverage will close the current gap with the {\it Fermi}-LAT band, its high energy coverage 
will open a new window on the sky and help us understand the intrinsic shape of the hardest blazar spectra. 

We outline the current status of CTA and discuss the science case for AGN physics with the observatory. 
Predictions for source detections based on extrapolations of {\it Fermi}-LAT spectra are discussed. 
An overview is given of prospects for the detection of extended emission from radio galaxies, of rapid
variability, and spectral features. The observation of AGN with CTA will also improve current constraints on the distribution 
of the extragalactic background light, the strength of the intergalactic magnetic field and Lorentz invariance 
violation.
\end{abstract}

\maketitle

\thispagestyle{fancy}


\section{Introduction}

The rapid increase --- in quantity and quality --- of scientific results from Imaging Air Cherenkov Telescope (IACT) arrays, such as H.E.S.S., MAGIC and
VERITAS, over the last ten years, has now firmly established the field of very high energy (VHE; defined here as E $\gtrsim$ 10 GeV) astrophysics. 
The rapid decrease of the photon spectra from astrophysical sources at high energies makes it unfeasible to observe $\gamma$-ray emission with space telescopes above a 
few 100 GeV, due to the limited collection area. Ground-based instruments are used instead, either in the form of air shower detectors such as e.g. HAWC, ARGO-YBJ or AS Gamma,
with large sky coverage, but usually very high energy thresholds or low sensitivity, or in the form of IACT arrays. The latter are arrays of telescopes optimized to detect the Cherenkov 
light flashes, in the ultra-violet and optical bands, from charged particles in air showers that develop when $\gamma$-rays or cosmic rays of sufficiently high energies 
enter the atmosphere.

In the VHE range, the sky is populated only with non-thermal sources, providing a direct view of galactic and extra-galactic sites of acceleration and interaction of
highly relativistic particles. At present, the detected sources include pulsar wind nebulae, supernova remnants, stellar clusters, molecular clouds,  and binary systems in our galaxy, 
and beyond it mostly Active Galactic Nuclei (AGN) and, up to now, two starburst galaxies. The main interest of VHE astrophysics is to reveal the conditions and mechanisms inside these sources 
that are responsible for the detected emission, and the underlying particle acceleration processes.

The next generation IACT array, currently in its Preparatory Phase, is the Cherenkov Telescope Array (CTA) \cite{cta_web}. In the first part of this
paper, a short update of the current status of the project will be given, while the second part focuses on the science case for AGN physics with CTA. 

\section{CTA in its Preparatory Phase}

\subsection{Timeline of the CTA project}

The Design Study for CTA started in 2007 and was financed mostly by individual institutes. The Conceptual Design Report that emerged from this
study \cite{cdr} provides the general science case for CTA and sets the framework for the array configurations and telescope designs. The project originated from the H.E.S.S. and MAGIC 
collaborations, but has today broad support from many astro(particle)-physicists and other interested scientists. At the time of writing, the international 
CTA consortium counts more than 800 scientists from 25 countries. Their common goal is to build the first open VHE observatory, with a significantly improved performance compared to the currently operating experiments. 

Listed among the priority projects on the roadmaps of several European networks (ESFRI, ASPERA/ApPEC, ASTRONET), CTA has received funding from the FP7 
programme for the Preparatory Phase. This phase has started in 2010 and will last three years. Among others, the detailed technical design of the instrument, array layouts, legal framework, data management and 
operation of the observatory are currently being established. The beginning of constructions for CTA is foreseen in 2014, allowing partial operation thereafter. Full operation of the arrays is planned for
2019. 

In 2010, the competing US project, AGIS~\cite{agis}, joined the CTA collaboration. It is planned that their support will provide an additional 36
telescopes to the southern array by 2021.

\subsection{Planned Performance and Design}

\begin{figure}[h]
\centering
\includegraphics[width=\columnwidth]{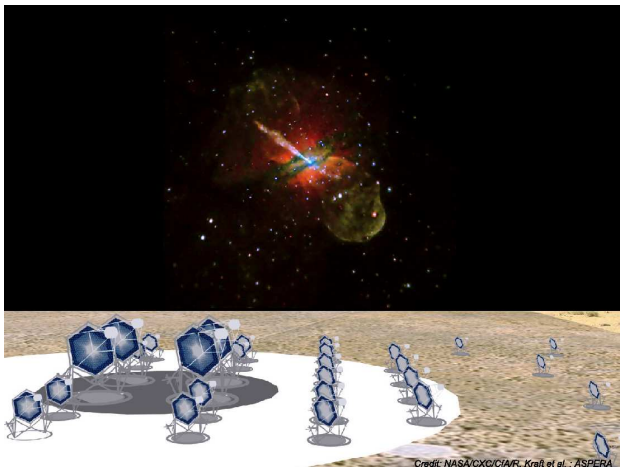}
\caption{Artist's view of the Cherenkov Telescope Array.} 
\label{fig:array}
\end{figure}

CTA is designed to bring about a similar leap forward for VHE astrophysics as is currently witnessed in high energy astrophysics thanks to the {\it Fermi} space telescope.
Its sensitivity will provide a gain by a factor of 10 compared to currently operating arrays, allowing the detection of point sources at the milli-Crab\footnote{The
flux from the Crab nebula (``one Crab''), about 2.26 $\times$ 10$^{-11}$ cm$^{-2}$ s$^{-1}$ when integrated above 1 TeV, is used as a general reference level in VHE astrophysics.} level for 50 hr of observation time at energies of a few TeV.
This will be achieved with the use of a much larger number of telescopes (several tens) than for current arrays, distributed over a large surface on the ground (of the order 
of 4 km$^2$) thus increasing the effective area of the array.

Energy coverage will also be extended, connecting to the {\it Fermi}-LAT range at the low end (about 30 GeV) while providing good coverage up to at least 100 TeV, 
thus opening a new window on the highest energies. This wide coverage will be possible for an acceptable cost due to the combination of differently sized telescopes (Fig.~\ref{fig:array}).
A few large size telescopes (LSTs) with a diameter of 24 m, situated in the center of the array, will be able to collect a sufficient number of Cherenkov photons even from weak air showers stemming from 
$\gamma$-rays at several 10 GeV, while a large number of small size telescopes (SSTs) with a diameter of 4 to 7 m, spread out over a large surface, will provide sufficient
effective area to collect bright but rare air showers at the highest energies. Intermediate energies will be covered by medium size telescopes (MSTs) with a diameter of 12 m.

The baseline optical system of the telescopes follows ``Davies-Cotton'' or parabolic designs, used by all of the currently operating arrays, but an alternative option, using secondary ``Schwarzschild-Couder'' optics,
is also being explored. Secondary optics might be implemented for the SST component and for an additional component of telescopes with a 10 m diameter primary mirror under study by the US
groups. The development of such systems is part of currently ongoing R\&D programs within the CTA consortium. The latter will further improve the sensitivity of the southern array by a factor of 2 to 4, as well as 
the angular resolution, given the denser arrangement of telescopes.

Massive simulations of $\gamma$-ray and cosmic-ray induced air showers and of the response of different array configurations are carried out, on the EGEE~\cite{egee} computing grid
and on computing centers at different institutes, to provide realistic estimates of the performance of CTA and to optimize its design with regard to the science requirements 
for a given cost.

Since VHE $\gamma$-rays are detected indirectly, the multiplicity of telescopes that detect a single air shower plays an important role in the quality of its reconstruction.   
An angular resolution down to the arc-minute will be achieved for bright events with energies of typically a few tens of TeV. A good astrometry, foreseen to be about 7 arc seconds, is 
essential to avoid source confusion, especially among Galactic sources, and to identify the emission region in extended sources. Energy resolution will be at the 10\% level to allow better resolution 
of spectral features than with the current instruments.

To fully exploit the information from the sky in the VHE range and to allow meaningful population studies, especially of extragalactic sources, CTA will have sites in the Northern 
and Southern Hemispheres, granting full sky coverage. The southern site will serve observations of galactic and extragalactic sources and will most likely consist of an array of SSTs, 
MSTs and LSTs, located in Argentina, Namibia or South Africa. The northern site, mostly of interest for extragalactic sources, which in general do not have spectra extending above 
a few 10 TeV, might not include an SST component. Sites for the northern array have been proposed in Arizona, on the Canary Islands, in China, India and Mexico.

\subsection{An Open Observatory}

CTA will be the first open observatory in the VHE range. Different from the operation of the present IACT arrays, observation time will be guaranteed for the scientific community and a 
significant part of the programme of observations will be compiled based on peer-reviewed proposals, similar to other major astrophysical facilities. 
While the actual observations will be conducted by a dedicated team of operators, data and tools for data reduction will be made available to the observer and the science 
community in general by a Science Data Center and will be accessible in a format compatible with standards of the Virtual Observatory. The elaboration of a specific data model for IACT 
data is among the tasks of the Preparatory Phase. Data handling and dissemination are also expected to make use of existing grid infrastructures (EGEE, GEANT).  

\section{Very High Energy Emission from AGN}

\begin{figure}[h]
\centering
\includegraphics[width=\columnwidth]{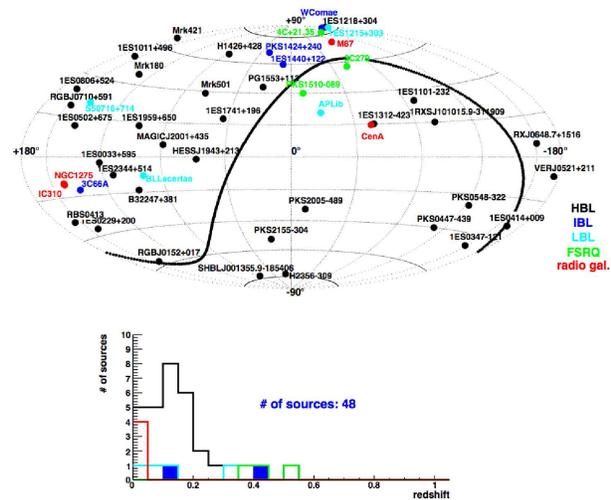}
\caption{AGN detected at VHE, status of November 2011 according to the TeVCAT catalog~\cite{tevcat}. The skymap is shown in the upper panel in galactic coordinates,
with the thick black line representing the celestial equator. The lower panel shows the distribution of redshifts. AGN from different classes are marked with different colours, 
as indicated in the legend.} 
\label{fig:skymap1111}
\end{figure}

Extragalactic sources make up about 40\% of the currently detected VHE sources.
With the exception of the two starburst galaxies NGC\,253 and M82, all extragalactic sources detected up to now at TeV energies
are AGN (Fig.~\ref{fig:skymap1111}). This selection of sources includes 4 radio-galaxies and 44 blazars (status as of November 2011). 
Blazars are radio-loud AGN with jets directly aligned with the line of sight, which substantially increases their luminosity due to Doppler boosting.
Their spectral energy distribution (SED) shows a bump in the optical to X-ray range, usually ascribed to synchrotron emission from a dense region
in the jet, and a second bump in the MeV to TeV range, interpreted either as Inverse Compton radiation from relativistic electrons in the emission region 
or by various processes involving relativistic hadrons. VHE emission from the radio-galaxies detected so far might be interpreted by considering these
sources as ``misaligned blazars'' (e.g.\ \cite{len2008}).

The large majority of the blazars detected at TeV energies today belongs to the class of high-frequency peaked BL Lacs (HBLs), featuring hard spectra at 
high energies and a peak of the synchrotron emission located in the UV/X-ray range. The redshift distribution of currently known VHE AGN, peaking above 0.1 
and extending up to $\sim$0.5, shows that most sources are detected relatively nearby (Fig.~\ref{fig:skymap1111}, lower panel). 

\section{The AGN Science Case for CTA}

A large fraction of the observation time of the southern array and most of the time of the northern array will be dedicated to AGN observations.
The VHE data from AGN provide access to the shortest variability time scales observed in these sources. They probe 
the high energy end of the underlying particle distribution and provide very useful insights into a variety of topics related to AGN physics, but also into the physics
of the extragalactic background light (EBL), intergalactic magnetic fields (IGMF) and fundamental physics. An in-depth discussion of the AGN science case
for CTA can be found in \cite{sol2012} (see also \cite{agnworkshop}), while in the following, an overview of selected topics is given.

\subsection{Fermi Extrapolations and Population Studies}

One domain where CTA will have a major impact will be population studies of different AGN classes at the highest energies. As shown above,
the current sample of VHE AGN is statistically very limited, especially for AGN classes other than HBLs. It is also biased towards strongly beamed, flaring, and
nearby sources. This makes it difficult to derive a VHE luminosity function or to use VHE data to improve the current understanding of AGN unification models 
or constraints on the EBL. The leap in sensitivity and the coverage of the Northern and Southern hemispheres provided by CTA will help to significantly increase the number of AGN detections, while its extension to
lower energies, where absorption on the EBL becomes negligible, will provide a much better reach in redshift.

One way to estimate the number of AGN accessible to CTA within a certain observation time is to extrapolate AGN spectra measured with {\it Fermi}-LAT to the VHE range 
and to compare them to the simulated sensitivity of CTA. We have extrapolated the power-law or log-parabolic spectra given in the 2FGL catalog \cite{2fgl} for the subset
of AGN in the ``clean'' sample and with known redshifts in the Roma BZCAT (3rd edition) \cite{bzcat} or Veron (13th edition) \cite{veron} catalogs. These spectra were then 
absorbed on the EBL, using the model by \cite{fran2008}, and an artificial break was introduced at 100 GeV only for the hardest sources with a photon index $\Gamma\le$2. The significance of 
the signal that would be observed with CTA for a given observation time was evaluated using simulated performance curves for a given array configuration (effective area, energy and 
angular resolution, background rate) and a detection was assumed for an excess significance of the signal over noise ratio of 5$\sigma$ (standard deviations), with at least 7 excess events 
and a signal exceeding 3\% of the background.

With only 0.5 hr of observation time per source, more than 25 sources should be detectable already. About 70, 170 and 230 sources would be detectable with a maximum of 5 hr, 50 hr and 
150 hr of observation time per source, respectively, which 
translates into total observation times of less than 2 months, about 3 years and less than 10 years, respectively. This is a rough, optimistic estimate, based on a total of 1500 hr per year available for AGN
observations, which would correspond for example to 500 hr for the southern array plus 1000 hr for the northern array. 

\begin{figure}[h]
\centering
\includegraphics[width=\columnwidth]{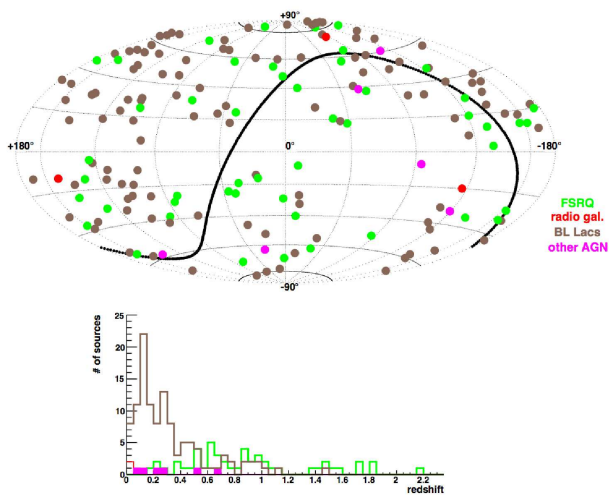}
\caption{Predicted AGN detections with CTA, based on extrapolations of {\it Fermi}-LAT spectra, for a maximum observation time of 50 hr per source. The skymap is shown in the upper panel 
in galactic coordinates. The lower panel shows the distribution of redshifts. AGN from different classes are marked with different colours, as indicated in the legend.} 
\label{fig:skymap50h_fermi}
\end{figure}

The resulting skymap and redshift distribution can be seen in Fig.~\ref{fig:skymap50h_fermi} for a maximum observation time per source of 50 hr.
It can be seen that CTA will be able to detect a significant fraction of the population of flat-spectrum radio quasars (FSRQs), which are mostly out of reach for current IACTs. A much larger 
range of redshifts will become accessible, with certain sources detectable out to redshifts of z$\sim$2.

\begin{figure}[h]
\centering
\includegraphics[width=\columnwidth]{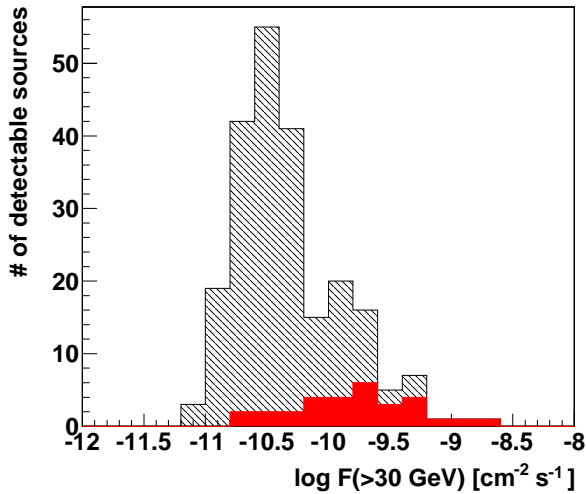}
\caption{The integrated fluxes above 30 GeV (in units of photons cm$^{-2}$s$^{-1}$) are shown based on extrapolations of {\it Fermi}-LAT spectra for a maximum exposure time per source of 150 hr
for all AGN detected by CTA (black, hatched histogram) and for those AGN detected by CTA that are already known as TeV emitters (red, filled histogram).} 
\label{fig:fluxes}
\end{figure}

The distribution of expected integrated fluxes from these extrapolations are shown in Fig.~\ref{fig:fluxes} for 150 hr maximum observation time per source. The lower flux limit of the distribution reflects the sensitivity of CTA. The comparison with the subset of 
already known TeV emitting AGN that are part of the selected sources from the 2FGL catalog (red filled histogram) emphasizes the gain in sensitivity that CTA will provide and its impact on future detections of TeV AGN.

Studies based on extrapolations from spectra at lower energies should however be taken with a pinch of salt. There is a risk of overestimating the number of detectable sources
since possible intrinsic spectral breaks above the {\it Fermi}-LAT band are ignored  (except for very hard spectra as described above) and would lead to lower fluxes in the CTA band.
In addition, the assumed EBL model might not provide a perfect description of the EBL density, which could lead to over- or under-estimations of the number of detectable sources at high 
redshifts. However, several recent discoveries of VHE AGN (e.g.\ PKS\, 0447-439, RGB\, J0648+152) were made based on predictions using similar methods.
In this particular study, we have also assumed that all sources are detected under a 20 degree zenith angle and with a certain array configuration that optimizes the low energy
coverage for the southern array. Several studies have been carried out based on similar extrapolations (e.g.\ \cite{has2012, ino2012}), showing that the number of detectable sources decreases 
by about 10\% when using a configuration for the southern or northern array that provides a more uniform sensitivity over the whole energy range and is therefore probably closer to the final array design. 
A significant loss of 30\% of detectable sources is predicted for configurations without an LST component, due to the important decrease of sensitivity at low energies.

On the other hand, several factors cause an underestimation of the number of detectable sources. The fact that only 34 out of the 45 VHE AGN detected up to the summer of 2011 appear also in the 
{\it Fermi}-LAT ``clean sample'' shows clearly that satellite-based and ground-based $\gamma$-ray telescopes do not probe exactly the same population of sources. An important fraction of BL Lacs with very 
hard spectra are not (or not clearly) detectable with {\it Fermi}-LAT. This can also be seen when comparing the simulated skymaps with Fig.~\ref{fig:skymap1111}. Several of the known VHE AGN do not appear 
in the simulated skymaps for this reason.
Neither do the estimations presented here account for the frequent flares or very active states of blazars that will push them over the
sensitivity threshold of CTA. One would also hope that new optical campaigns will provide redshift information for a larger fraction of known BL Lacs in the future, more than half of which are excluded
from the present study due to missing redshift information \cite{ack2011}. Finally, the expected addition of the US component to the southern array will further improve the sensitivity of CTA, which has not been taken
into account here. Given these considerations, the presented estimations seem rather conservative.

From the above it becomes apparent that the interest of CTA for AGN physics goes largely beyond a simple extension of the spectra of AGN detected with {\it Fermi}-LAT, even though
that in itself will already be an essential step towards constraining the underlying particle distribution and acceleration processes. CTA will not only complete the information on sources 
already detected with the LAT, but also detect sources not visible to the LAT. Furthermore, the high sensitivity of CTA will provide crucial information on flux and spectral variability on time scales not 
accessible to space telescopes.

Apart from targeted observations of known AGN (selected from the {\it Fermi}-LAT, X-ray or radio catalogs), unbiased population studies will most likely require sky surveys. Studies based
on a luminosity function \cite{ino2012} indicate that a ``wide and shallow'' coverage of the sky, investing only e.g.\ 0.5 hr per field of view with a full array, is the fastest way to 
maximize the number of detectable sources. A full sky survey could in this case be performed in less than one year, yielding however only around 50 sources, far less efficient than the follow-up of 
{\it Fermi}-LAT sources. A full sky survey with a deep coverage of 50 hr per field of view would exceed the planned 30 year lifetime of CTA and is thus not feasible, but a blank sky survey of part of the sky 
with relatively deep coverage is under consideration. A certain number of serendipitous detections of AGN, as already seen with present IACT arrays (e.g.\ \cite{hess_serendip, hil2011}), are certainly also 
expected. 

\begin{figure}[h]
\centering
\includegraphics[width=\columnwidth]{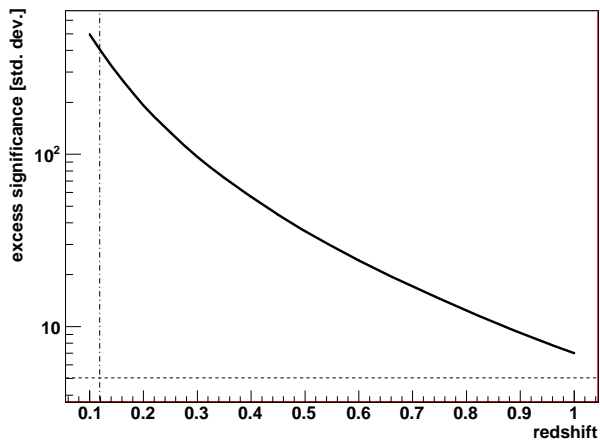}
\caption{Excess significance in standard deviations detected with CTA for a  simulated flaring source at different redshifts \cite{sol2012}. The characteristics of the source spectrum (flux level and spectral slope) have been adapted to the large VHE flares from PKS~2155-304 in 2006. A duration of 0.5 hr is assumed for the flaring state. The vertical dash-dotted line marks the redshift of PKS~2155-304. The horizontal dotted line indicates the 5$\sigma$ level.} 
\label{fig:flare_highz}
\end{figure}

The study of sources at high redshifts will be facilitated by the low energy reach of CTA, but also by the detection of flares. If one considers an extreme example like the flares detected from the very luminous HBL PKS\,2155-304 in 2006 \cite{pks2155_2006, pks2155_var}, at a redshift of z$=$0.116, one can evaluate the maximum redshift up to which very rapid and luminous flares would be detectable with CTA. 
Fig.~\ref{fig:flare_highz} shows the significance 
of the signal detected with CTA as a function of redshift for a flare with similar characteristics as the ones detected in 2006. Assuming a similar photon index and flux level and a duration of 0.5 hr, a comparable flare would 
be visible up to redshifts of z$\sim$1 for CTA, even for such a short duration. The detection of flaring states will thus further populate the source distribution at high redshifts.

Apart from increasing the number of VHE HBLs, CTA will also greatly improve the currently low statistics of FSRQs and of low and intermediate frequency peaked BL Lacs (LBLs, IBLs) detected at TeV energies. 
Its high sensitivity should also add a certain number of radio galaxies to the TeV source catalog. Predictions based on spectra from the {\it Fermi}-LAT might not be the best indicator to derive source counts for radio 
galaxies in the VHE range.
In the case of Centaurus A, for example, the VHE spectrum does not follow an extrapolation of the {\it Fermi}-LAT spectrum \cite{cena2010}. Other, similar cases might exist. It should also be noted that the radio galaxy
IC 310, detected at VHE, was not included in the predictions presented above, because it does not appear in the ``clean sample'' of {\it Fermi} sources.

There is a high probability that certain classes of AGN, not yet known to emit at TeV energies,  will be detected with CTA. A prime candidate
are Narrow Line Seyfert 1 galaxies (NLS1), about 7\% of which are thought to be radio-loud. Detection of several NLS1s with Fermi \cite{fermi_nls1}, including one in a flaring state \cite{fos2011}, promise good prospects for a detection
with CTA. These objects differ from other VHE AGN by the low masses of their central black holes and very high accretion rates. They thus offer information from a previously uncharted region in parameter
space, which could be of special interest for our understanding of the role that accretion plays in the disputed ``blazar sequence'' (for a recent review see e.g. \cite{geo2012}). 

The Seyfert 2 galaxy NGC\, 1068 has also been suggested as a potential VHE emitter \cite{len2010}, based on the detection of high energy emission with {\it Fermi}-LAT in  the direction of the source. If this emission can really be ascribed
to an outflow connected to the active nucleus, as suggested by the authors, rather than to starburst activity in the composite galaxy, this would present a first candidate of a radio-quiet AGN for VHE detection.
Such a discovery would be of great importance for the unification models of radio-loud and radio-quiet AGN.

\subsection{Extended Emission from Radio Galaxies}

The detection of radio galaxies in the VHE band raises the question of a possible detection of extended emission from these sources with CTA. Given the typical angular resolution of IACT arrays, this will
be a major challenge for even the most nearby sources, Centaurus A (at 3-5 Mpc) and M87 (at about 16 Mpc). Extended emission might in principle be seen from the kpc jets or from the large 
($\sim$100 kpc to Mpc scale) radio lobes of these objects and could teach us about their formation and energetics. 

Extended emission from the lobes of Centaurus A has been discovered with the {\it Fermi}-LAT \cite{cenalobes} and seems to be dominated by the Inverse Compton emission of electrons in the lobes
upscattering photons from the CMB  and EBL \cite{yan2012}. A simple extrapolation to the VHE band of the spectrum measured from the lobes does not predict a detectable signal for CTA. 
Somewhat more promising might be the emission from the extended jet. A simulation of the VHE emission from this jet \cite{har2011} leaves the possibility of detecting extended emission beyond 1 
arcminute  away from the central core with CTA. It should however be emphasized that an angular resolution of this order will only be reached for events with sufficiently high telescope multiplicities, i.e.\ for $\gamma$-rays of sufficiently high energies, typically not below 10 TeV. 

Resolving the kpc jet of M87, with an extension of roughly 30 arc seconds, is out of reach for CTA. But even if the jet will not be resolved, good astrometric precision might allow the detection of an offset between the peak 
of the VHE emission and the nominal source position of the radio core. In the case of M87 for example, the currently projected uncertainty in the absolute pointing might be sufficient to distinguish emission from the central core 
or from the ``knot A'' radio structure in the jet. A detection of emission from the radio lobes of M87 with CTA might be possible, if one assumes that 50\% of the total flux of the source is distributed over the lobes \cite{sol2012}. 
 
 Being able to pinpoint the VHE emission region, even for one or two radio galaxies, would present a milestone for AGN physics at the highest energies. The angular resolution and astrometry of CTA are therefore
 of real concern for AGN observations. For most of the detected AGN, a direct determination of the VHE emission region is obviously unfeasible, but the excellent sensitivity and energy resolution of CTA will help to characterize
 the emission region indirectly, through variability studies.

\subsection{Variability and the TeV-Jansky Connection}

Variability is frequently detected in blazars and radio galaxies and presents our best tool to constrain the size of the emission region of VHE $\gamma$-rays. Compared to other wavelength bands, this energy range shows the
most rapid variability, down to time-scales of a few minutes in the most extreme cases \cite{gai1996, pks2155_2006}. Given the usual light traveling time arguments, the measurement of the variability time scale $\Delta t_{obs}$ provides a direct constraint on the size of the emission region $R$ for a given Doppler factor $\delta$ : 
$R \, \delta^{-1} < c \, \Delta t_{obs}$/(1+z) .

In the case of the extreme flares from PKS\,2155-304 in 2006, more than 100 $\gamma$-rays per minute were detected with H.E.S.S., permitting
a measurement of time scales down to the minute. Flux doubling times of the order of two 
to three minutes were measured \cite{pks2155_2006}. A similar flare --- admittedly a rare event --- when observed with CTA, would yield thousands of $\gamma$-rays per minute, allowing an extension of the search for variability down to a time scale of several seconds \cite{bit2012}. Extremely small emission regions (or extremely high Doppler factors) might be found, further challenging our picture of the emission mechanisms. Alternatively, the detection of a break in the power spectral density would set a lower limit on the variability time scale. 

The variability of flaring sources will be studied in detail with data from CTA to characterize the underlying emission processes. A study carried out with H.E.S.S. data from the above mentioned flares found that the detected flux variability corresponds a lognormal process \cite{pks2155_var}, hinting at an underlying multiplicative mechanism. The implications of this observation are not yet fully understood and our
current perception would clearly profit from additional observational evidence. (See also \cite{bit2012, mch2012} for recent discussions of variability in blazars and expectations for CTA.)

Equally important as the study of flares will be the search for rapid variability during low states of the sources. It would help answer the question of whether the emission from blazars and radio galaxies is separated into a static, quiescent component to 
which flaring episodes are added, or whether variability is a general feature at all flux states. This requires a sufficiently high sensitivity to collect good photon statistics at low flux states. With CTA, this should be possible for the most luminous
blazars. PKS~2155-304, for example, should provide sufficient statistics for studies at the time scale of a few minutes, even in its low state.

If one wants to not only constrain the size of the emission region, but to pin down the origin of the VHE emission in the source, multiwavelength (MWL) campaigns of blazars and radio galaxies, especially during flares, are today the most promising means at our disposal. Several MWL campaigns on M87, for example, have shown that it is possible to infer the location of the emission region of VHE $\gamma$-rays by studying correlations between VHE and radio flares \cite{m87_1, m87_2}. This sort of study will certainly benefit greatly from the improvements in sensitivity and coverage of CTA and from the synergy with new and future radio astronomy facilities, such as e.g.  LOFAR and SKA. Furthermore, relations between the flux evolution in the VHE band and in the optical and radio bands, seen recently also in the data from PKS~2155-304 \cite{pks2155_mwl}, provide important insights into the emission processes and the radiative transfer inside the source.

\subsection{Emission Models and Spectral Features}

Good MWL coverage is crucial to provide sufficient information on the flux evolution and SED of the sources to constrain the many parameters of emission models and thus to identify the physical characteristics of the emission region and the underlying particle distribution. Such models are separated into leptonic and hadronic scenarios, depending on which type of particles dominate the emission at (very) high energies (for a recent review, see e.g.\ \cite{boe2012, rei2012} and
references therein). 

Leptonic models ascribe the high energy bump in the SED to Inverse Compton emission from electrons up-scattering synchrotron photons (synchrotron self-Compton, SSC) or photons from ambient fields (external Compton, EC). Attempts
are being made to interpret the ``blazar sequence'' with a variation of the EC contribution, considered negligible in HBLs and of major importance in FSRQs. Most data from VHE AGN available today are interpreted using stationary models, with the exception of the rare flaring events that provide sufficient statistics for time-dependent modeling of the flux and spectral variation. 

Careful multiwavelength modeling of flares provides valuable insights into the emission process. This can permit for example a discrimination of different components in the emission, as given for example by the SSC modeling of the 2006 flares from PKS\,2155-304 \cite{kat2008, pks2155_mwl}, where several emission regions with different characteristics were inferred from observed variability in the X-ray and VHE data. The higher sensitivity of CTA and the possibility of long-term monitoring of certain sources with a part of the array will increase the number of well-resolved flares. The adequate interpretation of these data will require, on the other hand, a higher level of sophistication of time-dependent emission models. 

\begin{figure}[h]
\centering
\includegraphics[width=\columnwidth]{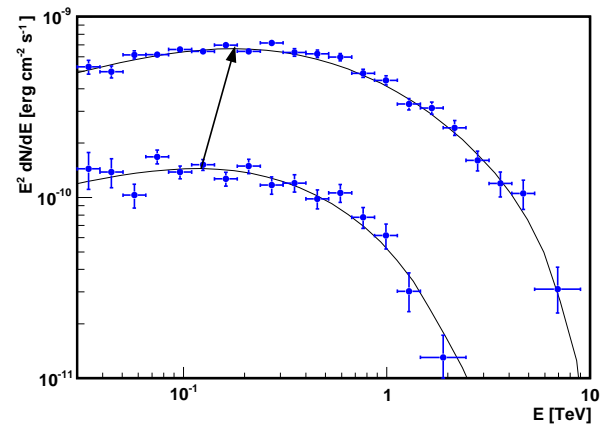}
\caption{Simulated spectral evolution as seen with CTA during the flare from Mrk421 detected on the night of March 18th, 2001, with Whipple. Model predictions based on a simple SSC scenario \cite{kat2012, sol2012} are shown for
two flux levels (thin lines), together with the spectra expected from CTA on a 15 minute time-scale (points with error bars). The movement of the SSC peak in this model is indicated by an arrow. For the given scenario, it would follow
the same path during the increase and decrease of the flare.} 
\label{fig:mrk421_ssc}
\end{figure}

As a further example, one can consider the TeV flare from Mrk 421, detected with Whipple in 2001 \cite{fos2008}. In the SSC framework, the flux variation during the flare can be explained with different scenarios, for example the injection of energetic particles, followed by radiative cooling, or the acceleration of particles accompanied by cooling, or simply a variation of the Doppler factor of the emission, which can be caused by geometric effects as the emission region propagates inside 
the jet. The modeled spectral evolution for this last case is shown in Fig.~\ref{fig:mrk421_ssc}, together with two simulated spectra as seen with CTA. The spectra have been simulated for a very short exposure of only 15 minutes. Their good
quality, even at this short duration, would permit a detailed time-dependent interpretation of the spectral change during the flare, which could distinguish between the different proposed scenarios \cite{sol2012}.

In hadronic models, a significant fraction of relativistic hadrons (usually assumed to be protons) is responsible for the emission in the high energy bump. The main processes that contribute to this emission are generally thought to be 
proton synchrotron radiation in strong magnetic fields and photo-pion production on internal and external photon fields. These processes generate photons and leptons at
very high energies, triggering pair-synchrotron or pair-Compton cascades in the source. Especially in FSRQs, which are characterized by important photon fields inside the source, emission from such cascades could lead to visible
features beyond the high energy end of the currently observed VHE spectra \cite{boe2009}. The detection of such a signature with CTA would provide evidence for hadronic emission in the source --- an issue of fundamental importance for the fields of ultra-high energy cosmic ray and astro-neutrino physics. 

A general problem with hadronic models are the longer time-scales expected for acceleration and cooling of hadrons, which make it very difficult to account for the observation of rapid flares in the VHE range. In this respect, it can be of 
interest to investigate ``lepto-hadronic'' models, where both Inverse Compton emission from the leptonic component and emission from the hadronic component contribute significantly to the high energy bump \cite{cer2011}, possibly
leaving detectable signatures in the (very) high energy spectrum.

In the spectra detected with {\it Fermi}-LAT from several FSRQs, some evidence has recently been found for spectral features from absorption of $\gamma$-rays on emission lines from the broad line region (BLR) \cite{pou2012}. 
A flux depression at the highest energies observed with the LAT (above a few GeV) in these sources can be interpreted as absorption on the Lyman recombination continuum of hydrogen and helium in the source. Some LBLs 
also seem to show absorption features in the LAT spectra. A promising candidate for example is the LBL S5 0617+714. A combination of the LAT and MAGIC spectrum from the source helps to better define the apparent absorption feature \cite{sen2011}.  
Such features should become clearly identifiable when combining spectra measured with {\it Fermi}-LAT and CTA. In the absence of simultaneous {\it Fermi} data, a flux depression at the low energy end of the CTA
spectrum might still provide some evidence for such features.  Absorption at higher energies, at a few 100 GeV, on the Balmer and Paschen lines might also be detectable with sufficiently good sensitivity and energy resolution\cite{pou2012}.

\subsection{Probing the EBL, IGMF and Lorentz Invariance with VHE $\gamma$-rays}
The improved capabilities of CTA for the observation of VHE emission from AGN will not only increase our knowledge of the sources themselves, but will also allow us to shed light on other topics, such as for example the distribution of 
the EBL or the strength of the IGMF. Detection of VHE $\gamma$-rays from AGN flares (or potentially from gamma-ray bursts), is also a means of probing Lorentz invariance violation (LIV). 

Constraints on the EBL level can be derived in different ways from VHE data from AGN (for a recent review, see e.g.\ \cite{maz2012}). Observations of the blazars H\,2356-309, 1ES\,1101-232, 1ES\,0229+200 with H.E.S.S. \cite{aha2006, aha2007} and 3C\,279 with MAGIC \cite{alb2008}, for example, provide an upper limit on the absorption by pair production on the EBL, under the assumption that the intrinsic photon index of the source in the VHE range is not harder than 1.5. The observed (absorbed) spectra were compared to the hardest acceptable intrinsic spectra under this assumption and an upper limit on the optical depth for pair production on the EBL was deduced, which is very close to the lower limit derived from direct measurements of the integrated light of resolved galaxies. 

The spectra that will be detected with CTA, with better photon statistics and a higher energy resolution, can be searched directly for imprints of the EBL absorption (e.g.\ \cite{orr2011, rau2010}). 
If one can exclude confusion with potential intrinsic signatures, this could be a way of directly measuring the EBL level and spectral form. The observation of luminous AGN over a large range of redshifts would allow a direct measurement of the evolution of the EBL density. 

With the advent of {\it Fermi}, another method of constraining the EBL has become possible, which makes use of the combination of the high energy and VHE spectra of a source. In the high energy range, i.e. below a few 10 GeV, 
absorption by the EBL can be neglected. The {\it Fermi}-LAT data thus present the intrinsic spectrum of the source, which can be extrapolated and compared to the absorbed VHE spectrum (e.g.\ \cite{orr2011, geo2010}). For standard emission
scenarios, an upper limit on the optical depth of the EBL can thus be derived based only on the assumption that no upturn occurs in the intrinsic spectrum in the VHE range. An exception to the validity of this assumption might be given in sources where the $\gamma$-ray flux suffers from absorption on BLR emission lines in the high energy range, while recovering in the VHE range, but this can be verified with the {\it Fermi}-LAT data. A problem in the interpretation of the
upper limit would be given by emission models for which absorption on the EBL is partially avoided (e.g. \cite{ess2011, pro2012, mur2012, dea2011}).

This method can also be reversed to obtain an upper limit on the redshift of a source \cite{fermi_z, zec2011}, when assuming a certain EBL model (see also \cite{pra2010}).  As stated above, a large fraction of those BL Lacs included in the 2FGL catalog have an unknown redshift, thus this information, together with lower limits from photometric methods, can help to provide at least a range for the redshifts of those sources that will be detected with CTA.

Given the frequent variability of VHE AGN, simultaneous data from {\it Fermi}-LAT and from CTA would be needed to guarantee that high energy and VHE data correspond to the same state of the source, if these methods were to profit from the higher quality spectra that will be provided by CTA.

The interaction of VHE $\gamma$-rays with photons from the EBL does not only lead to a suppression of the VHE flux, but it can result in detectable signatures at (very) high energies due to Inverse Compton up-scattering of photons from the cosmological microwave background by the produced electron-positron pairs. Depending on the strength of the IGMF, this can result in ``pair-echos'' or ``pair halos'' (for a recent review see \cite{inos2012} and references therein). The former, expected for weak fields below about 10$^{-16}$ G, are a delayed signal from the direction of the source that could be detected in the high energy range. For a very weak field and thus very weak diffusion of the produced pairs, a large fraction of the ``pair echo'' is expected to arrive from the direction of the source and to contribute to the high energy flux. By comparing {\it Fermi}-LAT spectra with spectra from Cherenkov telescopes of bright sources, lower limits can be put on the strength of the IGMF (e.g.\ \cite{ner2010, der2011, tay2011, tak2011}), with fundamental implications for our understanding of the formation of magnetic fields in the early Universe. 

Continuous data in the GeV and TeV range are needed to arrive at strong constraints. Simultaneous data from {\it Fermi}-LAT would be required to study the IGMF with this method with the superior capabilities of CTA. 

For stronger magnetic fields, the same process can lead to diffuse emission that would appear as a halo around the source and could be detected in the VHE range. Depending on the redshift of the source and the strength of the IGMF, this diffuse halo can extend to several degrees \cite{sol2012}. The large field of view of CTA will make it an ideal instrument to search for such a signal.

Last but not least, rapid flares from AGN at TeV energies can also be used to probe photon propagation and to set limits on LIV, predicted by certain quantum gravity theories. A detectable consequence of LIV would be delays in the photon
arrival times from distant sources due to the dispersion of light in the vacuum. Current limits are derived either from observations of GeV $\gamma$-rays from gamma-ray bursts \cite{fermi_grb} or from TeV $\gamma$-rays from AGN flares \cite{pks2155_liv}, since the induced time delay of the dispersion becomes larger with increasing photon energy and distance of the source. Current limits, if confirmed, would effectively rule out linear dispersion, thus falsifying certain quantum gravity theories, but quadratic dispersion terms, included in other theories, are not yet significantly constrained. Its access to blazars at high redshifts and the possibility to monitor sources to detect a maximum number of flares make CTA an ideal instrument to improve limits on LIV.

\section{Conclusions}

CTA will have a major impact on our knowledge of the astrophysics of non-thermal sources. Important implications are also expected for other domains, such as astroparticle physics, fundamental physics and cosmology. 
With regard to AGN physics, CTA will be able to provide for the first time statistically meaningful population studies of VHE sources, which will complete our knowledge on topics such as the AGN luminosity function or unification models.
Variability studies of AGN flares will be possible down to the scale of a few seconds for the most prominent events, providing strong constraints on size and location of the emission region. Probing rapid variability during low states will also become a possibility for some AGN.

The larger number of detectable AGN of different classes and the better photon statistics from individual sources will help to put much stronger constraints on emission models and to search for different spectral components or absorption
features. Information from AGN detected with CTA will also be used in several ways to further constrain the EBL density and the strength of the IGMF, with direct repercussions on open questions in cosmology. Another topic, not discussed here, is the measurement of a diffuse VHE flux, which, though technically difficult due to the high cosmic-ray background, would be of strong interest for the study of the fraction of unresolved sources and the search for dark matter signatures. 

Apart from a guaranteed science return, CTA provides an opportunity to discover new types of VHE sources, including new VHE AGN. Its high sensitivity up to at least 100 TeV will also broaden further our view of the universe.
The full potential of CTA will only be reached in MWL observations, including instruments over the whole electromagnetic spectrum. A synergy with detectors of astro-neutrinos, cosmic rays and gravitational waves is also anticipated.

\bigskip 
\begin{acknowledgments}
We gratefully acknowledge support from the agencies and organisations listed in this page: http://www.cta-observatory.org/?q=node/22 . 
\\ \\
The tools provided by D. Mazin were used to calculate the instrumental response of CTA from simulated performance curves. Figure 5 is 
based on work by M. Cerruti and B. Thiam. 
\end{acknowledgments}

\bigskip 

{\footnotesize


}

\end{document}